\documentclass[aps,twocolumn]{revtex4}
\usepackage{epsfig}

\begin{document}

\title{Stochastic dynamics of coupled systems and
spreading of damage}

\author{T. Tom\'e$^1$, E. Arashiro$^2$,
J. R. Drugowich de Fel\'{\i}cio$^2$, M. J. de Oliveira$^1$}

\affiliation{$^1$Instituto de F\'{\i}sica,
Universidade de S\~{a}o Paulo, \\
Caixa Postal 66318,
05315-970 S\~{a}o Paulo, S\~{a}o Paulo, Brazil \\
$^2$Departamento de F\'{\i}sica e Matem\'atica, FFCLRP, 
Universidade de S\~ao Paulo, \\
Av. Bandeirantes, 3900,
014040-901 Ribeir\~ao Preto, S\~ao Paulo, Brazil}

\date{\today}

\begin{abstract}
We study the spreading of damage in the one-dimensional
Ising model by means of the stochastic dynamics
resulting from coupling the system and its replica by
a family of algorithms that interpolate
between the heat bath and the Hinrichsen-Domany
algorithms. At high temperatures the dynamics is exactly mapped
into de Domany-Kinzel probabilistic cellular automaton.
Using a mean-field approximation and Monte Carlo
simulations we find the critical line that
separates the phase where the damage spreads 
and the one where it does not.

PACS numbers:
\end{abstract}

\maketitle

\section{Introduction}

It is well known that most models studied in 
equilibrium statistical mechanics, such as
the Ising model, are defined in a static way through
the equilibrium Gibbs probability distribution associated to the Hamiltonian
of the model. It is desirable from the theoretical and
numerical point of view to assign a dynamics to such models. The
stochastic dynamics introduced by Glauber \cite{glauber}
is the prototype example of a dynamics assigned to a 
static-defined model. The numerous versions of the
Monte Carlo method \cite{binder}, used in statistical mechanics
are also examples of dynamics assigned to static-defined models.
All of them are markovian processes that have the Gibbs 
probability distribution as the stationary distribution.
In general they are either a continuous time process 
governed by a master equation 
\cite{kaw,kampen,lig,self,marro,livrotm} 
or a probabilistic cellular automaton 
\cite{livrotm,dk,derrida,lebowitz,tome,gueu}. 
The latter is defined by a stochastic matrix, whose
elements are the transition probabilities, and the former by
the evolution matrix, whose nondiagonal elements are the
transition rates. 

If we wish to simulate, for instance, the Ising model we 
have to choose one of the possible stochastic dynamics since there 
are many. Having decided which dynamics to use,
that is, having decided which probabilistic rules to
use, we realize that there are several ways of doing the actual
simulation corresponding to the chosen probabilistic rules.
For instance, for the case of the
probabilistic cellular automaton used by 
Derrida and Weisbuch \cite{derrida}
to simulate the Ising model, and which will
concern us here, there are several ways of realizing 
the dynamics. We may use
the so called heat-bath algorithm \cite{barber} or the
algorithm introduced more recently
by Hinrichsen and Domany \cite{hd} or any other we
may invent. These algorithms govern the movement of the system in phase
space and they may be called {\it stochastic equations of motion} in phase
space. Different algorithms may be the
realization of the same probabilistic rule or stochastic dynamics.

The description of a system
either by the equation of motion or by the time evolution of
the probability are equivalent. 
An analogy can be made with the Brownian motion which
can be described either by the Langevin equation or by its 
associated Fokker-Planck equation \cite{kampen,livrotm,stoch,mjo}. 
The first is a stochastic equation of motion of a representative point 
in phase space whereas the second governs 
the time evolution of the probability distribution in phase space.

In the study of spreading of damage 
\cite{derrida,hd,grass1,creutz,stanley,mariz,grass2,tome,hwd,ara}
it has been realized that
algorithms that are realization of the same probalistic
rules may yield different results for the spreading of damage
\cite{mariz,hd,grass2,ara}, and they usually do.
The spreading of damage is a procedure through which we may 
study the sensibility of the time evolution of systems 
with respect to the initial conditions.
The procedure amounts to couple  
the system with a replica of it, 
each of them following the same equation of motion. 
The coupling is acomplished by the use of the
same sequence of random numbers. The equation of motion for each system
together with the use of the same random number define the equation of
motion for the coupled system from which we obtain the
{\it joint transition probability} \cite{tome,gueu}
for the coupled system. 

Suppose one uses an algorithm to couple 
a system and its replica.
This will lead us to certain joint transition probability. 
If another algorithm is used,
which is also a realization of the same transition
probability for a single system, the joint transition
probability will be distinct. The correlation between
system and replica will also be distinct and, in particular,
the Hamming distance which is a measure of the
damage spreading will be different.
For example, in one-dimensional Ising model, 
the heat-bath (HB) algorithm \cite{barber}
will give no spreading of damage whereas the 
Hinrichsen-Domany (HD) algorithm \cite{hd} will exhibit a spreading of
damage above a certain temperature \cite{hd}. 
This is an expressive example that the
spreading of damage is not a intrinsic static property of a given system,
but depends on the algorithm, or the stochastic equation of motion,
we use to perform the actual simulation \cite{hd,grass2}.

In this paper we introduce a family of algorithms, or
equations of motion, spanned by a
parameter that interpolates between the HB and HD
algorithms. The associated transition probability 
corresponds, for all values of the parameter, to the
Derrida-Weisbush (DW) probabilistic cellular automaton
\cite{derrida}.
If we use this family of algorithms to study the spreading
of damage, as we will do here,
the parameter will have no effect on each system separately 
since for any possible value of the parameter
the algorithm is related to the same transition probability.
However, the joint transition probability will depend on the parameter 
and the properties of the system, including the damage
spreading, will also depend on the parameter.
 
A remarkable property of the dynamics introduced here
is that at infinite temperature it is
exactly mapped into the Domany-Kinzel (DK) probabilistic cellular 
automaton \cite{dk}.
This gives support to a conjecture by Grassberger 
\cite{grass2} according to
which the generic class of damage spreading transitions
is the same as the directed percolation to which belong 
the transition ocurring in the DK probabilistic 
cellular automaton.

\section{Single system}

Let us consider a one dimensional
lattice where at each site one attaches an Ising variable 
$\sigma _i$ that takes the values $+1$ or $-1$ and denote by 
$\sigma=\{\sigma_i\}$ the set of all variables of the lattice.
The time evolution of the probability $P_{\ell}(\sigma)$
of state $\sigma$ at discrete time $\ell$ is given by
\begin{equation}
P_{\ell+1}(\sigma^{\prime})=
\sum_{\sigma} W(\sigma^{\prime}|\sigma)P_{\ell}(\sigma)
\end{equation}
where $W(\sigma^{\prime}|\sigma)$
is the transition probability from state $\sigma$ to state
$\sigma^{\prime}$ which, for a probabilistic cellular
automaton is given by \cite{livrotm}
\begin{equation}
W(\sigma^{\prime}|\sigma)=
\prod_i w_{PCA}(\sigma_i^{\prime}|\sigma)
\end{equation}
where $w_{PCA}(\sigma_i^{\prime}|\sigma)$ is the
probability that site $i$ will be in state $\sigma_i^{\prime}$
in the next step given that the present state of the
system is $\sigma$.
The DW probabilistic cellular automaton
\cite{derrida}
for the one dimensional Ising model is defined by
\begin{equation}
w_{PCA}(\sigma_i^{\prime}|\sigma)=
w_{DW}(\sigma_i^{\prime }|\sigma_{i-1},\sigma_{i+1}) 
\end{equation}
with
\begin{equation}
w_{DW}(+1|\sigma_{i-1},\sigma_{i+1})=p_i(\sigma )
\label{2}
\end{equation}
and 
\begin{equation}
w_{DW}(-1|\sigma_{i-1},\sigma_{i+1})=1-p_i(\sigma )
\label{3}
\end{equation}
where 
\begin{equation}
p_i(\sigma)=\frac{e^{-\beta J(\sigma _{i-1}+\sigma_{i+1})}}
{e^{\beta J(\sigma _{i-1}+\sigma_{i+1})}+
e^{-\beta J(\sigma _{i-1}+\sigma_{i+1})}} 
\end{equation}

The site $i$ assumes the state $+1$ with a probability
$p_i(\sigma)$ that does not depend on the central site 
$i$. If we choose the linear size of the system to be even
the dynamics is decomposed into two independent
dynamics for each sublattice.
It is possible to show \cite{derrida} that 
the DW probabilistic cellular automaton 
has as the stationary probability distribution
the Gibbs probability distribuiton
associated to the Ising model, namely, 
\begin{equation}
P(\sigma )=\frac 1Z\exp \{\beta J\sum_i\sigma _i\sigma _{i+1}\} 
\end{equation}
where $\beta=1/k_B T$,
so that it defines a stochastic dynamics that can be assigned
to the Ising model.

The transition probabilities 
$w_{DW}(\sigma_i^{\prime }|\sigma_{i-1},\sigma_{i+1})$
are shown in Table I 
where we used the parameter $p$ defined by
\begin{equation}
p=\frac{e^{-2\beta J}}
{e^{2\beta J}+e^{-2\beta J}}
\label{8a} 
\end{equation}

\begin{table}
\begin{center}
\caption{Transition probabilities for the DW
probabilistic celular automaton}
\begin{tabular}{|l|l|l|}
\hline
$w_{DW}$ & $+$ & $-$ \\ 
\hline
$++$ & $1-p$      & $p$      \\ 
$+-$ & $\frac 12$ & $\frac 12$ \\ 
$-+$ & $\frac 12$ & $\frac 12$ \\ 
$--$ & $p$        & $1-p$      \\
\hline
\end{tabular}
\end{center}
\end{table}

The actual computer realization of a probabilistic cellular 
automaton can be made in several ways. Here, we introduce a 
family of algorithms that are possible realizations
of the DW probabilistic cellular
automaton. It has a free parameter $a$ that interpolates
between the HD and HB algorithms.
At each time step all sites of the
lattice are updated in a synchronous way 
by means of the following algorithm, or
equation of motion for the spin variables, 
\begin{equation}
\sigma _i^{\prime }={\rm sign}\{p_i(\sigma )-\xi _i\}
\label{8}
\end{equation}
if $\sigma_{i-1}=\sigma _{i+1}$ and
\begin{equation}
\sigma _i^{\prime }={\rm sign}\{(a-\xi_i)(1-a-\xi_i)
(\frac 12-\xi _i)\}
\label{9}
\end{equation}
if $\sigma _{i-1}\neq \sigma _{i+1}$
where $\xi _i$ is a random number identically
distributed in the interval $[0,1]$.

When $a=0$ one recovers the HD algorithm \cite{hd} 
\begin{equation}
\sigma _i^{\prime }={\rm sign}\{p_i(\sigma )-\xi_i\} \qquad \sigma
_{i-1}=\sigma _{i+1} 
\end{equation}
\begin{equation}
\sigma _i^{\prime }=-{\rm sign}\{\frac 12-\xi_i\} \qquad \sigma
_{i-1}\neq \sigma _{i+1} 
\end{equation}
and when $a=1/2$ one recovers the HB algorithm \cite{barber,hd}
\begin{equation}
\sigma _i^{\prime }={\rm sign}\{p_i(\sigma )-\xi_i\} 
\end{equation}
It is straightforward to show that the algorithm defined 
by Eqs. (\ref{8}) and (\ref{9}) yields the
one-site transition probability
given by Eqs. (\ref{2}) and (\ref{3})
for any value of the parameter $a$.

\section{Coupled system}

Let us denote by $\sigma=\{\sigma_i\}$ and
$\tau=\{\tau_i\}$ the configurations of the system and its
replica, respectively. All sites of the system and its replica
are updated in a synchronous way according to the algorithm
\begin{equation}
\sigma _i^{\prime }={\rm sign}\{p_i(\sigma )-\xi_i\}
\label{13}
\end{equation}
if $\sigma_{i-1}=\sigma _{i+1}$ and 
\begin{equation}
\sigma _i^{\prime }={\rm sign}\{(a-\xi_i)(1-a-\xi _i)
(\frac 12-\xi_i)\}
\label{14}
\end{equation}
if $\sigma _{i-1}\neq \sigma_{i+1}$ and 
\begin{equation}
\tau _i^{\prime }={\rm sign}\{p_i(\tau )-\xi _i\}
\label{15}
\end{equation}
if $\tau_{i-1}=\tau_{i+1}$ and
\begin{equation}
\tau _i^{\prime }={\rm sign}\{(a-\xi_i)(1-a-\xi_i)
(\frac 12-\xi_i)\}
\label{16}
\end{equation}
if $\tau_{i-1}\neq \tau_{i+1}$. 
Notice that the random number $\xi _i$ is the same for both systems.

The coupled system will be described by a four-state probabilistic 
cellular automaton defined by the time evolution
\begin{equation}
P_{\ell +1}(\sigma^{\prime};\tau^{\prime})=\sum_{\sigma} \sum_{\tau}
W(\sigma^{\prime};\tau^{\prime}|\sigma;\tau)
P_{\ell}(\sigma;\tau)
\end{equation}
of the joint probability $P_{\ell}(\sigma;\tau)$ of state
$(\sigma;\tau)$ at discrete time $\ell$ where
$W(\sigma^{\prime};\tau^{\prime}|\sigma;\tau)$
is the joint transition probability from state 
$(\sigma;\tau)$ to $(\sigma^{\prime};\tau^{\prime})$,
and given by
\begin{equation}
W(\sigma^{\prime};\tau^{\prime}|\sigma;\tau)=\prod_i
w(\sigma _i^{\prime };\tau _i^{\prime }|
\sigma _{i-i},\sigma _{i+1};\tau_{i-1},\tau _{i+1})
\end{equation}
>From the stochastic equation of motion given by Eqs.
(\ref{13}), (\ref{14}), (\ref{15}), and (\ref{16}), 
we deduce the joint transition probabilities 
$w(\sigma _i^{\prime };\tau _i^{\prime }|
\sigma _{i-i},\sigma _{i+1};\tau_{i-1},\tau _{i+1})$ 
that the site $i$ of the system and the
replica assume the values $\sigma _i^{\prime }$ and $\tau _i^{\prime }$,
respectively.
The resultant joint transition probabilities
are displayed in Table II 
and are valid for $0\leq a\leq p$. 
For $p < a\leq 1$, the algorithm yields 
a joint transition probability which is independent of $a$ and is the
one that results by formally replacing, in Table II, $a$ by $p$.
The joint transition probability fulfill the following properties 
\begin{equation}
\sum_{\tau _i^{\prime }}w(\sigma _i^{\prime };\tau _i^{\prime }|
\sigma_{i-i},\sigma _{i+1};\tau _{i-1},\tau _{i+1})=
w_{DW}(\sigma _i^{\prime }|\sigma_{i-i},\sigma _{i+1})
\label{20a} 
\end{equation}
\begin{equation}
\sum_{\sigma _i^{\prime }}w(\sigma _i^{\prime };\tau _i^{\prime }|
\sigma_{i-i},\sigma _{i+1};\tau _{i-1},\tau _{i+1})=
w_{DW}(\tau _i^{\prime }|\tau_{i-1},\tau _{i+1})
\label{21a} 
\end{equation}
which contemplates the condition that the system 
and the replica follow
their own dynamics independent of the coupling.

The joint transition probabilities satisfy also 
the following properties. (a) Reflection symmetry
in which the states of sites $i-1$ and $i+1$ are interchanged, 
that is, $\sigma_{i-1} \leftrightarrow \sigma_{i+1}$ and
$\tau_{i-1} \leftrightarrow \tau_{i+1}$.
(b) System-replica symmetry in which the states of the
system and the replica are interchanged, that is, 
$\sigma_i \leftrightarrow \tau_i$ for all sites.
(c) Up-down symmetry defined by the transformation 
$\sigma_i \leftrightarrow -\sigma_i$ and
$\tau_i   \leftrightarrow -\tau_i$ for all sites.

\begin{table}
\begin{center}
\caption{Joint transition probabilities for the coupled system}
\begin{tabular}{|l|l|l|l|l|}
\hline
$w$ & $+;+$ & $+;-$ & $-;+$ & $-;-$ \\ 
\hline
$++;++$ & $1-p$            & $0$          & $0$       & $p$      \\ 
$+-;+-$ & $\frac 12$       & $0$          & $0$       & $\frac 12$ \\ 
$-+;-+$ & $\frac 12$       & $0$          & $0$       & $\frac 12$ \\ 
$--;--$ & $p$              & $0$          & $0$       & $1-p$      \\ 
$++;--$ & $p$              & $1-2p$       & $0$       & $p$      \\ 
$--;++$ & $p$              & $0$          & $1-2p$    & $p$      \\ 
$+-;-+$ & $\frac 12$       & $0$          & $0$       & $\frac 12$ \\ 
$-+;+-$ & $\frac 12$       & $0$          & $0$       & $\frac 12$ \\ 
$+-;++$ & $\frac 12-p+a$   & $p-a$        & $\frac 12-a$ & $a$ \\ 
$-+;++$ & $\frac 12-p+a$   & $p-a$        & $\frac 12-a$ & $a$ \\
$++;+-$ & $\frac 12-p+a$   & $\frac 12-a$ & $p-a$     & $a$ \\ 
$++;-+$ & $\frac 12-p+a$   & $\frac 12-a$ & $p-a$     & $a$ \\
$-+;--$ & $a$   & $\frac 12-a$ & $p-a$ & $\frac 12-p+a$ \\ 
$+-;--$ & $a$   & $\frac 12-a$ & $p-a$ & $\frac 12-p+a$ \\
$--;-+$ & $a$   & $p-a$ & $\frac 12-a$ & $\frac 12-p+a$ \\
$--;+-$ & $a$   & $p-a$ & $\frac 12-a$ & $\frac 12-p+a$ \\
\hline
\end{tabular}
\end{center}
\end{table}

The Hamming distance,
that characterizes the spreading of damage, is defined by
\begin{equation}
\Psi=\frac12 \langle 1-\sigma_i \tau_i \rangle
\end{equation}
which is also the order parameter related to the damage spreading
phase transition.

\section{Relation with the DK automaton}

In this section we show an exact relation
between the stochastic dynamics defined in Section III
and the DK probabilistic cellular
automaton \cite{dk}. 
If we let $\eta_i$ be the occupation variable
attached to site $i$, that is, $\eta_i=0$ or $1$ according to
whether site $i$ is empty or occupied by one particle, 
then the transition
probabilities
$w_{DK}(\eta^{\prime}_i|\eta_{i-1},\eta_{i+1})$
of the DK cellular automaton is given by
\begin{equation}
w_{DK}(1|00)=0 
\end{equation}
\begin{equation}
w_{DK}(1|01)=w_{DK}(1|10)=p_1 
\end{equation}
\begin{equation}
w_{DK}(1|11)=p_2 
\end{equation}
The DK cellular automaton displays a critical line
in the phase diagram $p_1$ versus $p_2$ that separates
the absorbing state, for which the density of particles is zero,
and the active state, for which the density is nonzero.

Now, let us denote by $\eta_i$ the coupling variable 
associated to the dynamics of Section III
that takes the value $1$ or $0$
according whether $\sigma _i\neq \tau _i$ or $\sigma _i=\tau _i$
respectively, given by 
\begin{equation}
\eta _i=\frac 12(1-\sigma _i\tau _i) 
\end{equation}
The relation between the Hamming distance
and the coupling variables is just
\begin{equation}
\Psi=\langle \eta_i \rangle
\label{45}
\end{equation}
The joint transition probabilities in the variables $\eta _i$
and $\sigma_i$ are defined by
\[
\tilde{w}(\sigma _i^{\prime };\eta _i^{\prime }|
\sigma _{i-i},\sigma _{i+1};\eta_{i-1},\eta _{i+1})=
\]
\begin{equation}
=w(\sigma _i^{\prime };\tau _i^{\prime }|
\sigma_{i-i},\sigma _{i+1};\tau _{i-1},\tau _{i+1}) 
\end{equation}
where $\tau_i=\sigma _i(1-2\eta _i)$

Summing over the coupling variable we get the following property 
\begin{equation}
\sum_{\eta _i^{\prime }}\tilde{w}(\sigma _i^{\prime };\eta _i^{\prime}|
\sigma_{i-i},\sigma _{i+1};\eta _{i-1},\eta _{i+1})=
w_{DW}(\sigma _i^{\prime }|\sigma
_{i-i},\sigma _{i+1}) 
\end{equation}
which contemplates, as in Eq. (\ref{20a}),
the condition that the system follows 
its own dynamics independent of the coupling.
The main property we wish to show, however, is that for 
infinite temperature, that is, for $p=1/2$ we have 
\begin{equation}
\sum_{\sigma _i^{\prime }}\tilde{w}(\sigma _i^{\prime };\eta _i^{\prime }|
\sigma_{i-i},\sigma _{i+1};\eta _{i-1},\eta _{i+1})
=w_{DK}(\eta _i^{\prime }|\eta_{i-i},\eta _{i+1}) 
\label{46}
\end{equation}
with the DK transition probabilites defined by 
$p_2=0$ and $p_1=1-2a$.
This means that the subsystem defined by the variables
$\{\eta_i\}$ follows a dynamics identical to
the DK probabilistic cellular automaton.
>From relation (\ref{45}) it follows that the Hamming distance coincides
with the order parameter of the active state 
displayed by the DK automaton. 

Yet for the case $p=1/2$, it is easy to show 
that the joint transition probability satisfies
the property 
\[
\tilde{w}(\sigma _i^{\prime };\eta _i^{\prime }|
\sigma _{i-i},\sigma _{i+1};\eta_{i-1},\eta _{i+1})=
\]
\begin{equation}
=w_{DW}(\sigma _i^{\prime }|\sigma _{i-i},\sigma_{i+1})
w_{DK}(\eta _i^{\prime }|\eta _{i-i},\eta _{i+1}) 
\end{equation}
with the DK transition probabilites defined by 
$p_2=0$ and $p_1=1-2a$. Therefore, the $\sigma$-subsystem
and the $\eta$-subsystem are statistically independent.

\section{Mean-field solution}

Dynamic mean-field approximation has already been used
to study systems in nonequilibirum stationary
states \cite{self,marro,tome,dickman}.
Here we set up equations for an approximate solution
of the equation that governs the time evolution of
the coupled system. We start by writing down
the equations that give the time evolution of the one-site 
and two-site probabilities, namely,
\[
P_{\ell+1} (\sigma_1;\tau_1)=
\sum_{\sigma_0,\sigma_2}\sum_{\tau_0,\tau_2}
w(\sigma_1;\tau_1|\sigma_0,\sigma_2;\tau_0,\tau_2)
\]
\begin{equation}
\times P_{\ell}(\sigma_0,\sigma_2;\tau_0,\tau_2) 
\label{20}
\end{equation}
and
\[
P_{\ell+1} (\sigma _1,\sigma _3;\tau _1,\tau _3)=
\sum_{\sigma _0,\sigma _2.\sigma _4}\sum_{\tau _0,\tau _2,\tau _4}
w(\sigma_1;\tau_1|\sigma_0,\sigma_2;\tau_0,\tau_2)
\]
\begin{equation}
\times w(\sigma_3;\tau_3|\sigma_2,\sigma_4;\tau_2,\tau_4) 
P_{\ell}(\sigma_0,\sigma_2,\sigma_4;\tau_0,\tau_2,\tau_4) 
\label{21}
\end{equation}
>From now on we will drop the subscript $\ell$ and use
the prime superscript for quantities calculated at time $\ell+1$.
To get a set of closed equations we use the approximation 
\[
P(\sigma_0,\sigma_2,\sigma_4;\tau_0,\tau_2,\tau_4)=
\]
\begin{equation}
=\frac 1{P(\sigma_2;\tau _2)}
P(\sigma_0,\sigma_2;\tau_0,\tau_2)
P(\sigma_2,\sigma_4;\tau_2,\tau_4) 
\end{equation}
which defines the dynamic mean-field 
pair approximation.

The probabilities $P(\sigma_1;\tau_1)$ and
$P(\sigma _1,\sigma _3;\tau _1,\tau _3)$ cannot
be considered all independent variables. 
Taking into account that they should have
the reflection symmetry and the system-replica
symmetry and, in addition, 
assuming the up-down symmetry the 
probabilities are related as follows
\begin{equation}
P(-;+)=P(+;-)=\frac12 \Psi
\end{equation}
\begin{equation}
P(-;-)=P(+;+)=\frac12 \Omega 
\end{equation}
\begin{equation}
P(--;--)=P(++;++)=A 
\end{equation}
\[
P(+-;--)=P(-+;--)=P(--;+-)=
\]
\[
=P(--;-+)=P(-+;++)=P(+-;++)=
\]
\begin{equation}
=P(++;-+)=P(++;+-)=B 
\end{equation}
\begin{equation}
P(-+;-+)=P(+-;+-)=C 
\end{equation}
\begin{equation}
P(--;++)=P(++;--)=D 
\end{equation}
\begin{equation}
P(-+;+-)=P(+-;-+)=E 
\end{equation}
These seven variables are not yet independent.
Only three of them can be considered independent which
we choose to be $\Psi$, $B$, and $D$.
The others are related to them by the
relations
\begin{equation}
\Omega=2P(+)-\Psi
\end{equation} 
\begin{equation}
A=P(++)-2B-D
\label{31}
\end{equation}
\begin{equation}
C=\frac12-P(++)-\frac12 \Psi+D
\label{32}
\end{equation}
\begin{equation}
E=\frac12 \Psi-2B-D 
\label{33}
\end{equation}
where $P(+)$ and $P(++)$ are the one-site and 
two-site probabilities corresponding to a 
single system. The exact solution of the one-dimensional
Ising model gives $P(+)=1/2$ and 
$P(++)=[1+(\tanh\beta J)^2]/4$. 

>From the time evolution given by
Eqs. (\ref{20}) and (\ref{21}) 
and using Eqs. (\ref{31}), (\ref{32}), and  (\ref{33})
we get the following closed equations for
$\Psi$, $D$, and $B$
\begin{equation}
\Psi^\prime =2\gamma D+8\alpha B
\label{34} 
\end{equation}
\[
D^\prime=4\alpha^2\frac {B^2}\Omega+
(4\alpha^2+\gamma^2) \frac {B^2}\Psi+
\]
\begin{equation}
+2\gamma (\gamma+2\alpha) \frac {D B}\Psi
+2\gamma ^2 \frac {D^2}\Psi
\label{35} 
\end{equation}
\begin{equation}
B^\prime=2\alpha B-4\alpha^2 \frac {B^2}\Omega
-4\alpha^2\frac {B^2}\Psi
-2\alpha\gamma \frac {DB}\Psi
\label{36} 
\end{equation}
where
\begin{equation}
\gamma =1-2p 
\end{equation}
and
\begin{equation}
\alpha =\frac12+p-2a 
\end{equation}

\begin{figure}
\epsfig{file=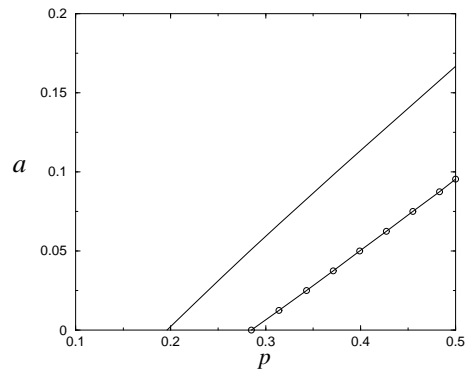,height=5cm}
\caption{Phase diagram in the plane $a$ versus $p$
where $p$ is relate to temperature by (\ref{8a}).
The continuous line corresponds to the
mean-field approximantion and the circles to the
Monte Carlo simulations.}
\end{figure}

A stationary solution of the evolution equation is such that
the stationary probability $P(\sigma;\tau)$ is 
zero when $\sigma \ne \tau$ which corresponds
to no damage spreading ($\Psi=0$).
>From Eqs. (\ref{34}), (\ref{35}), and (\ref{36})
we may obtain solutions with damage spreading ($\Psi \ne 0$).
The transition line is obtained by a linear analysis of
stability of the solution around $\Psi=0$ and by assuming 
that the variables $B$ and $D$ vanishes linearly with $\Psi$. 
Taking the limit $\Psi \to 0$
we find a transition line given by the implicit equation
\[
(1-\alpha)^2 \gamma^3-4\alpha(3\alpha^2-5\alpha+2) \gamma^2+
\]
\begin{equation}
+4\alpha^2(13\alpha^2-16\alpha+5)\gamma-8\alpha(3\alpha-2)=0
\end{equation}
whose solution is shown in the phase diagram of Fig. 1.
In particular, when $a=0$ (corresponding to the HD algorithm)
we have $\gamma=2(1-\alpha)$ which substituted
in the equation for the transition line gives
\begin{equation}
1-9\alpha+33\alpha^2-59\alpha^3+53\alpha^4-20\alpha^5=0
\end{equation}
whose solution is $\alpha=0.696173$ from which 
we get $p=0.196173$
so that $J/k_B T_c=0.352597$ and $T_c=2.83610$.
When $a=1/2$ (correspoding to the HB algorithm)
there is no transition.

At infinite temperature, $p=1/2$, the mean-field
transition line gives $a=1/6$. Now, using the relation
$p_1=1-2a$ obtained from the equivalence
with the DK automaton, and taking into
account the result $p_1=2/3$
obtained in \cite{tome} in the pair approximation for the DK automaton
we have $a=1/6$ in coincidence with
our present result.

\section{Numerical simulations}

Our numerical simulations resulted in the 
transition line shown in Fig. 1.
When $a=0$ we have obtained $p=0.285(1)$ which gives
$J/k_B T_c=0.230(1)$ and $T_c=4.35(2)$ in agreement with the
result by Hinrichsen and Domany \cite{hd}, namely $J/k_B T_c=0.2305$.
At infinite temperature, $p=1/2$, the numerical results
give a transition at $a=0.0955(1)$. Now, using the
relation $p_1=1-a$ obtained from the
mapping of our model into the 
DK cellular automaton, we obtain $p_{1c}=0.809(1)$ 
in agreement with previous Monte Carlo numerical results,
namely $p_{1c}=0.8095(5)$ \cite{rieger}.

\begin{figure}
\epsfig{file=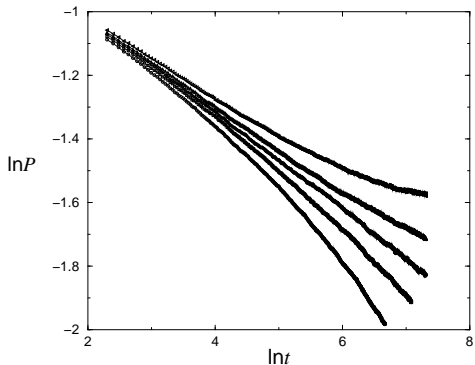,height=5cm}
\caption{Time dependent Monte Carlo simulations for the damage
survival probability $P$ for a lattice with linear size
$L=1000$. Numerical data are shown for 
$a=0.075$ and 
$p=0.450$, $0.453$, $0.455$, $0.457$, and $0.460$
from bottom to top.}
\end{figure}

The determination of the critical line
was obtained by using the time dependent
method  \cite{torre,marro,grass2}.
We started with two one-dimensional lattices
(system and replica) with $L=1000$ sites.
Both lattices were initialized with completly indepedendent
random configurations so that half the spins were
damaged at the begining ($\Psi=1/2$).
The update was done in a synchronized way by using
the algorithm defined
by Eqs. (\ref{13}), (\ref{14}), (\ref{15}), and (\ref{16}),
with the same random number for both lattices.
The damage surviving probabilites $P(t)$, 
obtained by taking the averages
over 2000 samples, were collected from $t=1$
to $t=1500$ Monte Carlo steps. 
At the critical point
we expect the following asymptotic time behavior
\begin{equation}
P(t) \sim t^{-\delta}
\end{equation}
Therefore, a double-log plot of $P$
versus $t$ will be linear at the critical point. 
In Fig. 2 we show how the critical value of
$p$ was found when $a=0.075$. 
Several values of $p$, the ones shown in Fig. 2, were checked
in order to find a linear behavior in a log-log
plot of $P(t)$ versus $t$. 
Our estimate in this case gives $p_c=0.455(1)$
for $a=0.075$. The straight line
fitted to the numerical data gives 
$\delta=0.16(1)$ in agreement
with a transition belonging to the
direct percolation universality class \cite{marro}.
For other values
of $a$ the procedure were the same.

\section{Conclusion}

We have introduced a family of algorithms 
to describe the time evolution of
the one-dimensional Ising model
The family of algorithms interpolates
between the HB and the HD algorithms
and the resulting stochastic dynamics 
corresponds to the DW probabilistic
cellular automaton.
Coupling a system with its replica by using
the same sequence of random numbers, we have
determined the joint transition probability
which defines a four-state probabilistic 
cellular automaton.
By using a dynamic pair mean-field approximation and
Monte Carlo simulations we have found that the stochastic
dynamics defined by the family of algorithms displays
a line of critical  points separating a phase where the
damage spreads and a phase where it does not. 
One important feature of the joint stochastic dynamics
studied here is that at infinite
temperature the joint dynamics is exactly mapped into
the DK probabilistic cellular automaton.
This result together with the Monte Carlo simulations 
give support to a conjecture by Grassberger according
to which the damage spreading transition is in the
universality class of the directed percolation.


\bigskip

\begin{references}

\bibitem{glauber} R. J. Glauber, J. Math. Phys. {\bf 4}, 294 (1963).

\bibitem{binder}{\it Monte Carlo Methods in Statistical Physics},
edited by K. Binder, 2nd. ed. (Springer-Verlag, Berlin, 1986).

\bibitem{kaw} K. Kawasaki in {\it Phase Transitions and
Critical Phenomena}, edited by C. Domb and M. S. Green
(Academic Press, New York, 1972), vol. 2, p. 443.

\bibitem{kampen} N. G. van Kampen, {\it Stochastic Process
in Physics and Chemistry} (North-Holland, Amsterdam, 1981).

\bibitem{lig} T. M. Liggett, {\it Interacting Particle Systems}
(Spinger-Verlag, New York, 1985).

\bibitem{self} T. Tom\'e and M. J. de Oliveira,
Phys. Rev. A {\bf 40}, 6643 (1989).

\bibitem{marro} J. Marro and R. Dickman, {\it Nonequilibrium
Phase Transition in Lattice Models} (Cambridge University Press,
Cambridge, 1999).

\bibitem{livrotm} T. Tom\'e e M. J. de Oliveira, 
{\it Din\^amica Estoc\'astica e Irreversibilidade}
(Editora da Universidade de S\~ao Paulo, S\~ao Paulo, 2001).

\bibitem{dk} E. Domany and W. Kinzel, Phys. Rev. Lett. {\bf 53}, 447 (1984).

\bibitem{derrida} B. Derrida and G. Weisbuch, Europhys. Lett. 
{\bf 4}, 657 (1987).

\bibitem{lebowitz} J. L. Lebowitz, C. Maes and E. R. Speer,
J. Stat. Phys. {\bf 59}, 117 (1990).

\bibitem{tome} T. Tom\'e, Physica A {\bf 212}, 99 (1994).

\bibitem{gueu} E. P. Gueuvoghlanian and T. Tom\'e,
Int. J. Mod. Phys. B {\bf 11}, 1245 (1997).

\bibitem{barber} M. N. Barber and B. Derrida, J. Stat. Phys.
{\bf 51}, 877 (1988).

\bibitem{hd} H. Hinrichsen and E. Domany, Phys. Rev. E {\bf 56}, 94 (1997).

\bibitem{stoch} T. Tom\'e and M. J. de Oliveira, 
Braz. J. Phys. {\bf 27}, 525 (1997).

\bibitem{mjo} M. J. de Oliveira, Int. J. Mod. Phys. B 
{\bf 10}, 1313 (1996).

\bibitem{grass1} P. Grassberger, Physica A {\bf 214}, 547 (1995).

\bibitem{creutz} M. Creutz, Ann. Phys. {\bf 167}, 62 (1986).

\bibitem{stanley} H. Stanley, D. Stauffer, J. Kertesz, and H. Hermann, 
Phys. Rev. Lett. {\bf 59}, 2326 (1987).

\bibitem{mariz} A. M. Mariz, H. J. Hermann, and L. de Arcangelis,
J. Stat. Phys. {\bf 59}, 1043 (1990).

\bibitem{grass2} P. Grassberger, J. Stat. Phys. {\bf 79}, 13 (1995).

\bibitem{hwd} H. Hinrichsen, J. S. Weitz and E. Domany,
J. Stat. Phys. {\bf 88}, 617 (1997).

\bibitem{ara} E. Arashiro and J. R. Drugowich de Fel\'{\i}cio,
Braz. J. Phys. {\bf 30}, 677 (2000).

\bibitem{martins} M. L. Martins, H. F. Verona de Rezende, C. Tsallis 
and A. C. N. de Magalh\~aes, Phys. Rev. Lett. {\bf 66}, 2045 (1991).

\bibitem{torre} P. Grassberger and A. de la Torre,
Ann. Phys. (NY) {\bf 122}, 373 (1979). 

\bibitem{dickman} R. Dickman, Phys. Rev. A {\bf 34}, 4246 (1986).

\bibitem{rieger} H. Rieger, A. Schadschneider and M. Schreckenberg,
J. Phys. A {\bf 27}, L423 (1994).

\end{references}
\end{document}